\documentclass{aa}

\usepackage[colorlinks, allcolors=blue]{hyperref}
\usepackage{amsmath} 

\usepackage{graphicx}
\usepackage{xcolor}
\usepackage{txfonts}
\usepackage{longtable}
\usepackage{geometry}
\usepackage{datetime}
\usepackage{lscape}
\usepackage{float}
\usepackage{url}
\usepackage{orcidlink}

\def\arcsec{$^{\prime\prime}$}

\newcommand{\tablecomments}[1]{
    \par\vskip1pt
    \noindent
    \vrule height 11pt depth 2pt width 0pt
    \texttt{Note} -- 
    #1\par
    \vskip1pt
}

\begin{document} 

\title{Detection of low-luminosity X-ray pulsations from the accreting millisecond pulsar IGR~J17511$-$3057: An ever-thinning thread between bright accretion and sub-luminous states}

\author{G.~Illiano\inst{1} \orcidlink{0000-0003-4795-7072}, 
 A.~Papitto\inst{2} \orcidlink{0000-0001-6289-7413}, 
 S.~Campana\inst{1} \orcidlink{0000-0001-6278-1576}, 
 A.~Marino\inst{3,4,5} \orcidlink{0000-0001-5674-4664},
 A.~Miraval Zanon\inst{6} \orcidlink{0000-0002-0943-4484}, 
 F.~Carotenuto\inst{2} \orcidlink{0000-0002-0426-3276},
 F.~Coti Zelati\inst{3,4,1} \orcidlink{0000-0001-7611-1581},
 M.~C.~Baglio\inst{1} \orcidlink{0000-0003-1285-4057}, 
 F.~Ambrosino\inst{2} \orcidlink{0000-0001-7915-996X},
 C.~Malacaria\inst{2} \orcidlink{0000-0002-0380-0041}, 
 C.~Ballocco\inst{2} \orcidlink{0009-0001-0155-7455}, 
 G.~K.~Jaisawal\inst{7} \orcidlink{0000-0002-6789-2723},
 M.~M.~Messa\inst{1,8},
 E.~Parent\inst{3,4} \orcidlink{0000-0002-0430-6504},
 T.~D.~Russell\inst{5} \orcidlink{0000-0002-7930-2276},
 A.~Sanna\inst{9} \orcidlink{0000-0002-0118-2649},
 A.~Tzioumis \inst{10} \orcidlink{0000-0002-0988-7969}
 }

 \institute{
 INAF–Osservatorio Astronomico di Brera, Via Bianchi 46, I-23807, Merate (LC), Italy\\
\email{giulia.illiano@inaf.it}
 \and
 INAF-Osservatorio Astronomico di Roma, Via Frascati 33, I-00078, Monte Porzio Catone (RM), Italy
 \and Institute of Space Sciences (ICE, CSIC), Campus UAB, Carrer de Can Magrans s/n, E-08193 Barcelona, Spain
 \and Institut d’Estudis Espacials de Catalunya (IEEC), E-08860 Castelldefels (Barcelona), Spain  
 \and INAF, Istituto di Astrofisica Spaziale e Fisica Cosmica, Via U. La Malfa 153, I-90146 Palermo, Italy
 \and ASI - Agenzia Spaziale Italiana, Via del Politecnico snc, 00133 Roma, Italy
 \and DTU Space, Technical University of Denmark, Elektrovej 327-328, DK-2800 Lyngby, Denmark
 \and Dipartimento di Fisica, Università degli Studi di Milano, Via Celoria 16, I-20133 Milan, Italy
 \and Dipartimento di Fisica, Università degli Studi di Cagliari, SP Monserrato-Sestu, KM 0.7, Monserrato, 09042 Italy
 \and Australia Telescope National Facility, CSIRO, PO Box 76, Epping, New South Wales 1710, Australia
 }             

\date{}
\authorrunning{Illiano et al.}
\titlerunning{Low-luminosity X-ray pulsations from IGR~J17511$-$3057}
 
\abstract{
After nearly a decade in quiescence, the accreting millisecond pulsar IGR~J17511$-$3057 displayed a new outburst on 2025 February 11, its third since discovery, following previous activity in 2009 and 2015. We report on an \textit{XMM-Newton} Target of Opportunity observation performed on 2025 March 4, more than twenty days after the outburst onset. From the X-ray spectrum -- well described by an absorbed Comptonization model -- we estimated an unabsorbed 0.5$-$10~keV luminosity of $L_X \sim 7 \times 10^{33} \, \mathrm{erg \, s^{-1}}$ (assuming a source distance equal to the upper limit of $6.9$~kpc). To put this into context, we analyzed an archival \textit{Chandra} observation performed in 2019, which yielded a quiescent luminosity of $L_\mathrm{X,q} \sim 2 \times 10^{32} \, \mathrm{erg \, s^{-1}}$ in the same energy band. Although this comparison indicates that the source was still well above its quiescent level during the \textit{XMM-Newton} observation, the estimated low luminosity during the late stage of the 2025 outburst would typically place the source in the propeller regime. Nevertheless, we unexpectedly detected coherent X-ray pulsations with an amplitude peaking at $\sim$42\% in the 0.3$-$3~keV band. We also observed a spectral softening compared to the early stages of the outburst. Finally, we report a 3$\sigma$ upper limit of 60\,$\mu$Jy\,beam$^{-1}$ on the source flux density at 5.5~GHz from ATCA observations acquired on 2025 April 12, following a decline in the accretion activity, as indicated by our analysis of \textit{NICER} data from 2025 March 15, which revealed no significant X-ray pulsations at a luminosity level of $L_X \sim 1 \times 10^{34} \, \mathrm{erg \, s^{-1}}$. We discuss our findings in the context of other accreting millisecond pulsars and draw comparisons with transitional systems in the sub-luminous disk state.
}


\maketitle
 
\section{Introduction} \label{sec:intro}
Accreting millisecond pulsars (AMSPs; see \citealt{DiSalvo_Sanna_2022ASSL, Patruno_Watts_2021ASSL, Campana_DiSalvo_2018ASSL} for reviews) are relatively low-magnetized ($B \sim 10^{8}-10^{9} \, \mathrm{G}$), rapidly rotating ($P_{\textrm{spin}} \lesssim 10 \, \mathrm{ms}$) neutron stars (NSs), typically found in tight binary systems with low-mass ($\lesssim 1 \, \mathrm{M_\odot}$) companions. They attain their rapid spin rates through a Gyr-long phase in which they shine as bright low-mass X-ray binaries (LMXBs), fueled by accreting matter and angular momentum transfer from the sub-solar donor star. When accretion stops, the NSs turn on as rotation-powered millisecond pulsars, making AMSPs their likely progenitors \citep{Alpar_1982Natur, Radhakrishnan_Srinivasan_1982CSci}. 

AMSPs predominantly persist in quiescence, where little or no mass accretion takes place (X-ray luminosity of $L_X \lesssim 10^{31}-10^{33} \, \mathrm{erg \, s^{-1}}$; see, e.g., \citealt{Campana_DiSalvo_2018ASSL}), but they can suddenly exhibit mass accretion outbursts lasting from weeks to months, which are marked by a considerable increase in the X-ray luminosity ($L_X \sim 10^{36}-10^{37} \, \mathrm{erg \, s^{-1}}$). During outbursts, the in-flowing matter from the companion star is funneled along the NS magnetic field lines and is accreted onto its magnetic poles, giving rise to coherent X-ray pulsations at the millisecond spin period that provide direct probes of the NS properties and emission physics.

The connection between accretion-powered and rotation-powered pulsars has been further strengthened by the discovery of transitional millisecond pulsars (tMSPs; see \citealt{Papitto_deMartino_2022ASSL} and references therein), which can swing between these two states on timescales of days, likely driven by changes in the mass transfer rate from the companion. In addition to these two main states, tMSPs exhibit an intermediate condition, dubbed the sub-luminous disk state, characterized by the presence of an accretion disk, an X-ray luminosity of $L_X \simeq 10^{33}-10^{34} \, \mathrm{erg \, s^{-1}}$, and a $\gamma$-ray luminosity up to ten times higher than that of typical rotation-powered millisecond pulsars (see, e.g., \citealt{Stappers_2014ApJ, Takata_2014ApJ}; see also \citealt{Torres_Li_2022ASSL} for a review). This intermediate configuration is particularly relevant to this work, as it provides a framework to investigate whether AMSPs, during the late phases of their outbursts, show similarities with tMSPs in the sub-luminous disk state.
A hallmark of this state is erratic X-ray emission, with rapid and unpredictable switches between two distinct intensity levels, named high and low modes \citep{Linares_2014}, alongside occasional flaring episodes \citep[see, e.g.,][]{Tendulkar_2014, Bogdanov_2015ApJ_J1023}. 
In the prototype PSR~J1023$+$0038 \citep{Archibald_2009Sci}, coherent pulsations have been detected simultaneously in X-ray, UV, and optical bands during X-ray high modes only, vanishing in low modes \citep{Archibald_2015, Ambrosino_2017, Papitto_2019ApJ, Jaodand2021, Miraval_Zanon_2022}. This finding, along with the similar pulse profiles and the pulsed spectral energy distribution compatible with a single power law \citep{Papitto_2019ApJ, Miraval_Zanon_2022}, suggests a common emission mechanism. Standard rotation- or accretion-powered mechanisms cannot individually explain all properties \citep{Ambrosino_2017, Papitto_2019ApJ}. Of particular relevance to this work is the fact that, although X-ray pulsations were initially linked to magnetically channeled accretion onto the NS poles \citep{Archibald_2015}, this process hardly explains the much higher-than-expected optical pulsed luminosity \citep{Ambrosino_2017, Papitto_2019ApJ}. This led to a scenario in which X-ray, UV, and optical pulsations during high modes are due to synchrotron radiation from a boundary region $\sim$100~km from the NS, where the relativistic wind from a rotation-powered pulsar interacts with inflowing matter from the inner accretion disk (\citealt{Papitto_2019ApJ, Veledina_2019}; see also \citealt{Illiano_2023AA, Baglio_CotiZelati_2023A&A, CotiZelati2024, Baglio_2025ApJ, Illiano_2025A&A} for recent observations supporting this scenario).

To date, IGR~J18245$-$2452 has been the only tMSP observed to undergo a full accretion outburst in 2013, reaching luminosities and durations typical of classical AMSPs \citep{Papitto_2013Natur}. Earlier episodes, more consistent with a sub-luminous disk state, were identified in 2008 and 2009 archival data \citep{Linares_2013MNRAS, Pallanca_2013}. While the 2008 \textit{Chandra} light curve showed clear mode-switching, no such behavior was detected in 2013 despite a similar luminosity. However, due to the shorter exposure and lower count rate, the authors could not rule out the presence of comparable variability. Spectral softening was also observed toward the end of the 2013 outburst, with the photon index increasing from $\Gamma \sim 1.3$ to $\Gamma \sim 2.5$ as the X-ray luminosity dropped by about two orders of magnitude \citep{Linares_2013MNRAS}.

The X-ray transient IGR~J17511$-$3057 was discovered during an $\sim$1-month-long outburst in September 2009 \citep{Baldovin_2009ATel, Bozzo_2010A&A, Markwardt_2009ATel_quiescence}. Soon after its discovery, the detection of 245 Hz X-ray pulsations confirmed its classification as an AMSP \citep{Markwardt_2009ATel}. The NS is in a 3.47-hour orbit \citep[e.g.,][]{Riggio_2011A&A} with a main-sequence companion star that has a mass between 0.15 and 0.44~$\mathrm{M_\odot}$ \citep{Papitto_2010MNRAS}. Following an $\sim$5.5-year-long quiescent period, a second outburst was detected in March 2015 \citep{Bozzo_2015ATel, Papitto_2016A&A}.  
Multiple type-I X-ray bursts were observed during both the 2009 and 2015 outbursts \citep[e.g.,][]{Altamirano_2010MNRAS, Bozzo_2010A&A, Papitto_2010MNRAS, Falanga_2011A&A, Papitto_2016A&A}, with many showing burst oscillations close to the NS spin frequency \citep{Altamirano_2010MNRAS, Papitto_2016A&A}. None of the bursts exhibited signs of photospheric radius expansion, allowing only for an upper limit on the source distance of 6.9~kpc \citep{Altamirano_2010MNRAS}.

After nearly a decade in quiescence, IGR~J17511$-$3057 showed a third outburst detected by \textit{INTEGRAL} on 2025 February 11 \citep{Sguera_2025ATel_new_outburst, Sguera_2025ATel17061_Integral_monitoring}, and was confirmed by \textit{NICER} through the detection of coherent X-ray pulsations at the known spin frequency \citep{Ng_2025ATel17032}. 
In this paper, we report on a \textit{XMM-Newton} Target of Opportunity (ToO) observation performed on 2025 March 4, when the source had already faded to a relatively low luminosity level.
We also report on a \textit{NICER} observation performed on 2025 March 15 and a radio observation with the Australia Telescope Compact Array (ATCA) on 2025 April 12. 
To put the low luminosity observed with \textit{XMM-Newton} into context, we analyzed an archival \textit{Chandra} observation from 2019, when the source was in quiescence.
In Sect.~\ref{sec:obs}, we describe \textit{XMM-Newton}, \textit{NICER}, \textit{Chandra}, and ATCA observations and data reduction. The X-ray spectral analysis of \textit{XMM-Newton}, \textit{NICER}, and \textit{Chandra} is presented in Sect.~\ref{sec:spectral_analysis}, while the timing analysis of \textit{XMM-Newton} and \textit{NICER} data is detailed in Sect.~\ref{sec:timing_analysis}. Finally, in Sect.~\ref{sec:discussion}, we provide an interpretation of our findings and discuss their implications, while Sect.~\ref{sec:conclusions} summarizes our main conclusions.

\section{Observations} \label{sec:obs}
\subsection{\textit{XMM-Newton}} \label{sec:XMM_data}
\begin{figure}
    \centering
    \includegraphics[width=0.47\textwidth]{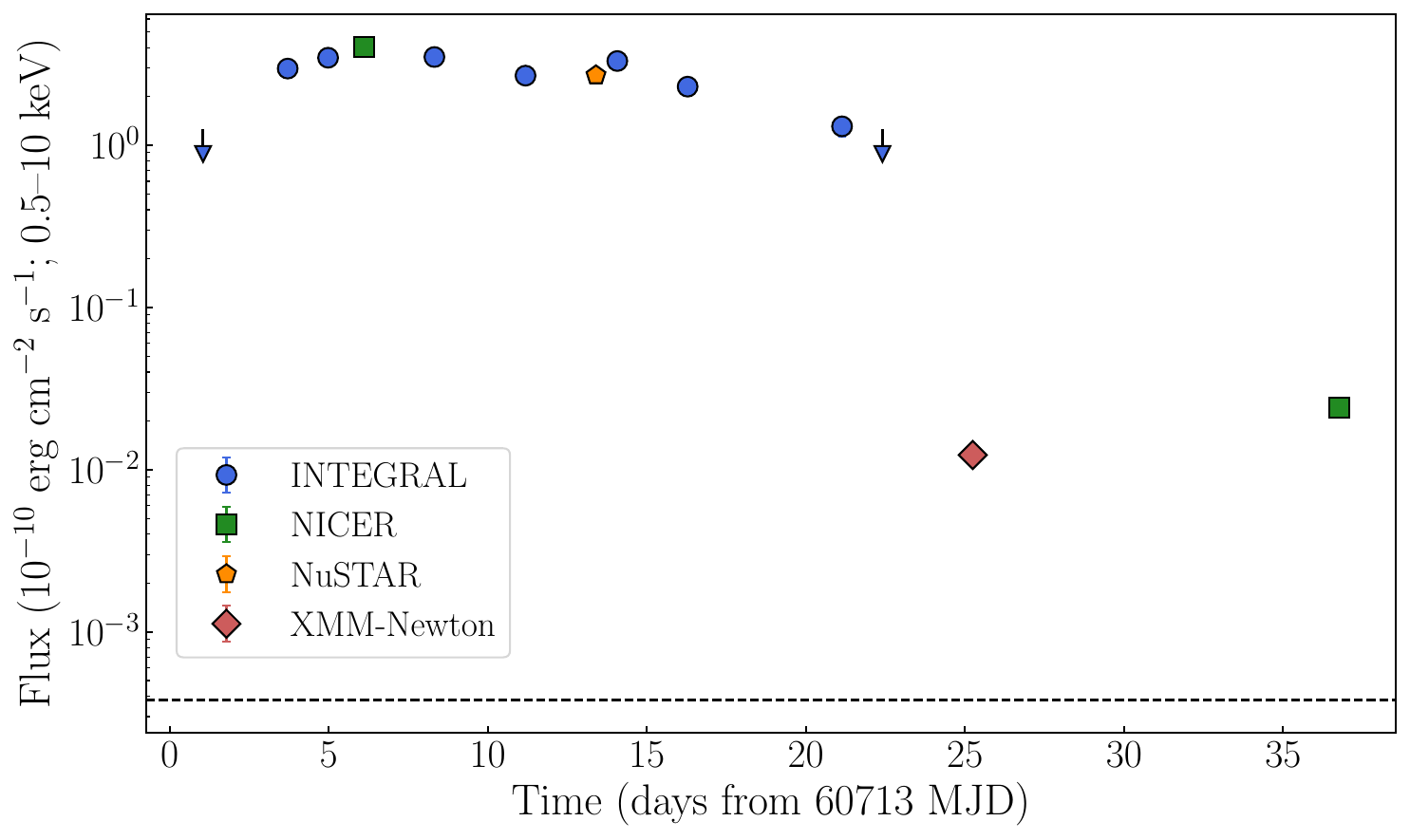}
    \caption{Light curve in the 0.5$-$10~keV band of the 2025 outburst of IGR~J17511$-$3057 observed by \textit{INTEGRAL} (blue points for detections, blue downward arrows for 3$\sigma$ upper limits; \citealt{Sguera_2025ATel17061_Integral_monitoring}), \textit{NICER} (green squares; the first point is from \citealt{Sanna_2025AA}, while the last point is discussed in Sect.~\ref{sec:NICER_spectral_analysis}), \textit{NuSTAR} (orange pentagon; \citealt{Sanna_2025AA}), and \textit{XMM-Newton} (red diamond). The times on the x-axis are given in days since 60713 MJD, corresponding to the first \textit{INTEGRAL} observation reported for this outburst \citep{Sguera_2025ATel17061_Integral_monitoring}.    
    The associated error bars are not visible, as they are smaller than the size of the data points. The dashed black line indicates the unabsorbed 0.5$-$10~keV flux of $3.8 \times 10^{-14} \, \mathrm{erg \, cm^{-2} \, s^{-1}}$ estimated during quiescence from the 2019 \textit{Chandra} observation (Sect.~\ref{sec:Chandra_analysis}). 
    The \textit{INTEGRAL} 28–60~keV fluxes, initially expressed in mCrab, were converted to the 0.5–10~keV band and reported in CGS units. We assumed the Crab pulsar’s spectral model from \citet{Kirsch_2005SPIE} (a power law with a photon index $\Gamma=2.252$ and a normalization $\mathrm{N}= 15.47$ absorbed by an equivalent hydrogen column of $\mathrm{N_H} =0.4 \times 10^{22} \, \mathrm{cm^{-2}}$, adopting the interstellar medium abundance and the cross-section tables from \citet{Wilms_2000ApJ} and \citet{Verner_1996ApJ}, respectively). The spectrum of IGR~J17511$-$3057 was described with an absorbed semiphenomenological Comptonization model with a seed blackbody component as reported in \citet{Sanna_2025AA} (see Model-I for \textit{NuSTAR}, Table 2).} \label{Fig:outburst_lc}
\end{figure}

\textit{XMM-Newton} \citep{Jansen_2001AA, Schartel_2022} observed IGR~J17511$-$3057 on 2025 March 4 (Obs. ID: 0954191401; PI: Illiano), $\sim$21 days after the onset of the outburst (see Fig.~\ref{Fig:outburst_lc} and Sect.~\ref{sec:XMM_spectral_analysis} for details), once the source became visible to the satellite following an initial period of visibility constraints. The EPIC-pn \citep{Struder_2001AA} operated in fast timing mode, offering a time resolution of 29.5~$\mathrm{\mu s}$, while the two EPIC-MOS cameras \citep{Turner_2001AA} were set to small window mode, providing a time resolution of 0.3~s. We processed and analyzed the Observation Data Files (ODF) using the Science Analysis Software (SAS; v.21.0.0).
We corrected the photon arrival times to the Solar System Barycenter using the JPL ephemerides DE405 \citep{Standish_DE405} and the \texttt{barycen} task. We adopted the source coordinates R.A. (J2000) $= 17^{\mathrm{h}}51^{\mathrm{m}}08.66^{\mathrm{s}}$ and DEC. (J2000) $=-30^\circ57' 41.0^{\prime\prime}$ \citep{Paizis_2012ApJ}.
We excluded the final $\sim$13.5, $\sim$15.7, and $\sim$15.6~ks of data from the EPIC-pn, MOS1, and MOS2, respectively, due to elevated background flaring activity in the 10$-$12~keV light curves. The resulting net exposures were $\sim$20~ks for each of the EPIC instruments.
For the EPIC-pn, source photons were extracted from a 11-pixel-wide strip centered on the brightest pixel column (RAWX=32-42), while background photons were taken from a strip with the same width far from the source position (RAWX=3-13). For each MOS, source photons were extracted from a circular region with a 40\arcsec\, radius centered on the source position, and background photons from a 80\arcsec\, wide, source-free circular region on one of the outer CCDs. No type-I X-ray bursts were detected during the \textit{XMM-Newton} observation.

\subsection{\textit{NICER}} \label{sec:NICER}
The Neutron star Interior Composition Explorer (\textrm{NICER}; \citealt{NICER_Gendreau_2012}) observed IGR~J17511$-$3057 during two separate intervals. The first set of observations, conducted between 2025 February 11 and 14, captured the onset of the X-ray outburst and are discussed in detail by \citet{Sanna_2025AA}. In this work, we focus on the observations performed on March 15, 2025 (Obs. ID: 7512010105), following a period during which the source was not visible. We analyzed the NICER data using HEASoft version 6.34 and NICERDAS version 12, with calibration files from CALDB version 20240206. Data processing was carried out using the \texttt{nicerl2} task. The obtained cleaned event files were used to generate spectral products via the \texttt{nicerl3-spect} task, with \texttt{SCORPEON} background model (version 23) with the setting \texttt{bkgformat=file}. 
Spectra were grouped using the \texttt{ftgrouppha} tool, automatically invoked by the \texttt{nicerl3-spect} task, applying the ``optimal binning'' scheme \citep{Kaastra_Bleeker_2016A&A}, with a minimum of 25 counts per energy bin (using the options \texttt{GROUPTYPE=OPTMIN} and \texttt{GROUPSCALE = 25}).
Using the \texttt{nicerl3-lc} pipeline, we extracted the 0.3$-$10~keV light curve, which showed an average count rate of $\sim$1~cts~s$^{-1}$, significantly lower than the rate of 52~cts~s$^{-1}$ observed at the beginning of the outburst \citep{Sanna_2025AA}. No type-I X-ray bursts were detected during this \textit{NICER} observation.
We corrected the photon arrival times to the Solar System Barycenter using the same ephemerides and source coordinates adopted for the \textit{XMM-Newton} data (see Sect.~\ref{sec:XMM_data}).

\subsection{\textit{Chandra}} \label{sec:Chandra_obs}
IGR~J17511$-$3057 was observed in quiescence by the \textit{Chandra} X-ray observatory \citep[e.g.,][]{Weisskopf_2000SPIE_Chandra} on 2019 July 27 (Obs. ID: 21228). The observation was performed using the Advanced CCD Imaging Spectrometer (ACIS; \citealt{Garmire_2003SPIE}), with the source placed on the back-illuminated S3 chip (CCD ID: 7) of the spectroscopic array (ACIS-S). The total exposure time was $\sim$29.6~ks, and data were acquired in faint mode.
We processed and analyzed the data using the \textit{Chandra} Interactive Analysis of Observations (\texttt{CIAO}, v.4.17; \citealt{Fruscione_2006SPIE}) software package, along with the calibration files from CALDB version 4.12.0. 
The dataset was first reprocessed with the \texttt{chandra\_repro} tool using default parameters. The source was detected using \texttt{wavdetect} with default parameters, yielding a net count rate of $\sim$$1.2 \times 10^{-3}$~counts~s$^{-1}$. 
Source and background spectra were extracted using the \texttt{specextract} tool. We grouped the spectra using the \texttt{GRPPHA} tool from the \texttt{FTOOLS} software package, ensuring at least one count per bin.

\subsection{ATCA}
We conducted radio observations of IGR~J17511$-$3057 on 2025 April 12 using the Australia Telescope Compact Array (ATCA) as part of program CX598 (PIs: Carotenuto \& Russell). To maximize \textit{uv}-coverage\footnote{In radio interferometry, \textit{uv}-coverage refers to the sampling of spatial frequencies by the projected baselines between antennas. As the Earth rotates, these projections change, tracing elliptical tracks in the \textit{uv}-plane and improving image reconstruction. See \citet{Thompson_2017isra} for more details.}, the observations were split into two segments: the first was performed from 17:19~UT to 18:25~UT, and the second from 22:53~UT to 23:59~UT. ATCA was undergoing a correlator upgrade, limiting data acquisition to a single band with a central frequency of 5.5~GHz and a 2-GHz bandwidth. For calibration, we used B1934$-$638 as the primary calibrator for both bandpass and flux density, while PKS 1741-312 served as the complex gain calibrator. The data were processed using standard flagging, calibration, and imaging techniques within the Common Astronomy Software Applications ({\tt CASA}) package \citep{casa2}. Imaging was performed with the \texttt{tclean} task, adopting a Briggs weighting scheme with a robust parameter of 0.5. After combining the two intervals, no radio counterpart was detected at the location of IGR~J17511$-$3057. 
Furthermore, due to the 10-second time resolution, these interferometric observations are not sensitive to the narrow, individual pulses typical of most radio MSPs, which exhibit low duty cycles and short pulse widths \citep[e.g.,][]{Manchester_2005AJ_ATNF}. We derived a 3$\sigma$ upper limit on the total flux density of IGR~J17511$-$3057 of 60\,$\mu$Jy\,beam$^{-1}$ at 5.5~GHz. 
This result is discussed in Sect.~\ref{sec:discussion_radio}.

\section{Analysis and results} 
\subsection{Spectral analysis} \label{sec:spectral_analysis}
We performed the X-ray spectral analysis on \textit{Chandra} archival data acquired during quiescence in 2019 (Sect.~\ref{sec:Chandra_analysis}), as well as on \textit{XMM-Newton} (Sect.~\ref{sec:XMM_spectral_analysis}) and \textit{NICER} data (Sect.~\ref{sec:NICER_spectral_analysis}) obtained during the late stage of the 2025 outburst. All spectra were fit using the X-ray spectral fitting package \texttt{XSPEC} \citep{Arnaud_1996_XSPEC} version 12.14.1. We adopted the interstellar medium abundance and the cross-section tables from \citet{Wilms_2000ApJ} and \citet{Verner_1996ApJ}, respectively. All uncertainties on spectral parameters are given at the 1$\sigma$ confidence level.

\subsubsection{\textit{XMM-Newton}} \label{sec:XMM_spectral_analysis}
The average EPIC-pn spectrum was extracted with a minimum of 50 counts in each channel. Data below 0.7~keV were excluded due to calibration issues in the EPIC-pn data acquired in timing mode\footnote{\url{https://xmmweb.esac.esa.int/docs/documents/CAL-TN-0018.pdf}.}. The final analysis was restricted to the 1$-$7~keV band, as the background dominates outside this range. The average EPIC-MOS1 and EPIC-MOS2 spectra were extracted in the 0.3$-$10~keV band, with a minimum of 25 counts in each channel. We included a renormalization factor in our spectral modeling to address cross-calibration uncertainties between the different instruments and always verified that these factors were consistent within 10\%. 

\begin{figure}
    \centering
    \includegraphics[width=0.47\textwidth]{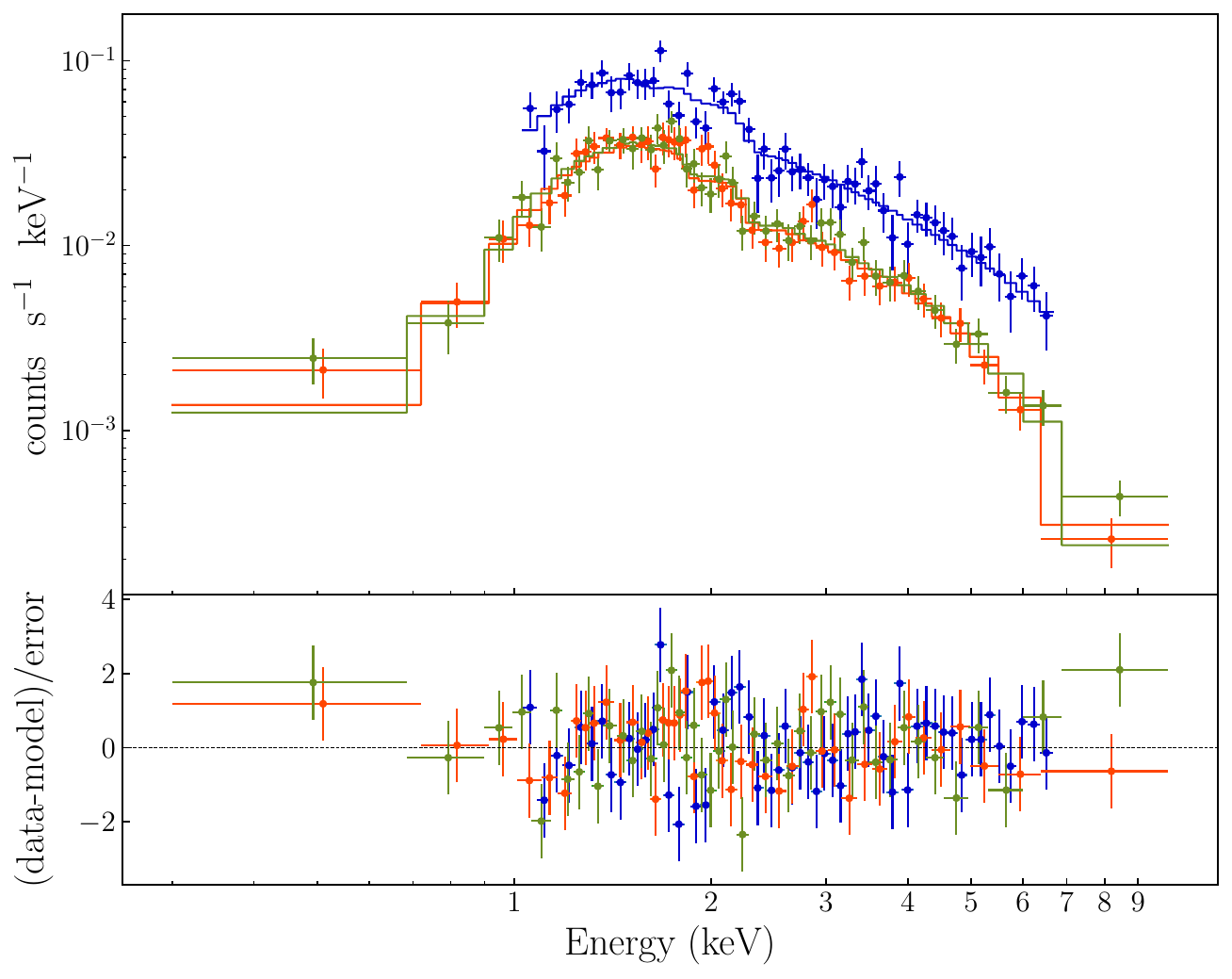}
    \caption{X-ray spectrum of IGR~J17511-3057 as observed by \textit{XMM-Newton} on 2025 March 4 with EPIC-pn (blue), EPIC-MOS1 (orange), and EPIC-MOS2 (green), along with the best-fitting model \texttt{constant * TBabs * (thComp*bbodyrad)} (see Table~\ref{tab:params_spectrum}). The bottom panel displays the residuals from the fit.} \label{Fig:total_spectrum}
\end{figure}

To better constrain other parameters, we fixed the absorption column density to $N_\mathrm{H} = 1.06 \times 10^{22} \, \mathrm{cm^{-2}}$, as obtained by \citet{Sanna_2025AA}. We use this value consistently throughout the rest of the paper.
We first fit the spectrum with an absorbed power-law model (\texttt{constant * TBabs * powerlaw}), yielding a $\chi^2$ = 169.88 for 149 degrees of freedom (d.o.f. in the following). Since the X-ray spectra of AMSPs in outbursts are typically described by thermally Comptonized emission \citep[e.g.,][]{DiSalvo_2023hxga.book}, we replaced the power-law component with the convolution model \texttt{thcomp} \citep{Zdziarski_2020MNRAS} applied to the \texttt{bbodyrad} component (\texttt{constant * TBabs * (thcomp * bbodyrad)}; see Fig.~\ref{Fig:total_spectrum}). When left as a free parameter, the covering fraction, representing the fraction of Comptonized photons, always reached its hard limit set at 1, and was therefore fixed at this value. Due to the lack of spectral coverage above 10~keV, we fixed the electron temperature, $\mathrm{kT_e}$, to 20~keV based on the results of \citet{Sanna_2025AA}. 
This revised model provided a significantly improved fit, with a decrease in $\chi^2$ of about 35 for one fewer degree of freedom compared to the power-law model.
The best-fit parameters are listed in Table~\ref{tab:params_spectrum}. The unabsorbed 0.5$-$10~keV flux, $F_\mathrm{{0.5-10}}$, was estimated by including the \texttt{cflux} component.

\begin{table} 
\renewcommand{\arraystretch}{1.2}
\centering
\caption{Best-fit spectral parameters from the \textit{XMM-Newton} observation for the model \texttt{constant * TBabs * (thComp*bbodyrad)}.} \label{tab:params_spectrum}
\begin{tabular}{l c c}          
   \hline\hline
    Component & Parameter & Value \\
    \hline
    \scshape{Tbabs} & $\mathrm{N_H}$ ($10^{22}$ cm$^{-2}$) & $1.06^{(*)}$\\
    \scshape{thcomp} & $\Gamma$ & $2.20^{+0.07}_{-0.06}$ \\
    & $\mathrm{kT_e}$ (keV) & $20^{(*)}$ \\
    & $\mathrm{cov\_frac}$ & $1^{(*)}$\\
    \scshape{bbodyrad} & $\mathrm{kT}$ (keV) & $0.33 \pm 0.02$ \\
    & $\mathrm{Norm_{bbodyrad}}$ & $6^{+2}_{-1}$\\
    \scshape{cflux} &$F_\mathrm{{0.5-10}} \, \mathrm{(10^{-12} \, erg \, cm^{-2} \, s^{-1})}$ & $1.23 \pm 0.04$\\
    \hline
    &$\chi ^2$/d.o.f & 134.51/148\\
    \hline
\end{tabular}
\tablecomments{
$\Gamma$ is the photon index, $\mathrm{kT_e}$ is the electron temperature, $\mathrm{cov\_frac}$ is the covering fraction, and $\mathrm{Norm_{bbodyrad}}$ is the normalization for the \texttt{bbodyard} component.
$F_\mathrm{{0.5-10}}$ is the unabsorbed 0.5$-$10~keV flux.
$^{(*)}$ Kept frozen during the fit.}
\end{table}
\noindent

Figure~\ref{Fig:outburst_lc} compares the estimated unabsorbed 0.5–10 keV flux from Table~\ref{tab:params_spectrum} with values obtained by \textit{INTEGRAL} \citep{Sguera_2025ATel17061_Integral_monitoring}, \textit{NICER}, and \textit{NuSTAR} \citep{Sanna_2025AA} during the 2025 outburst of IGR~J17511$-$3057. Our flux estimate is about two orders of magnitude lower than those reported earlier in the outburst. Using the upper limit on the source distance of $6.9 \, \mathrm{kpc}$ -- estimated by \citet{Altamirano_2010MNRAS} from the absence of photospheric radius expansion in any of the detected type-I X-ray bursts -- we derived an X-ray luminosity in the 0.5$-$10~keV band of $L_X \sim 7 \times 10^{33} \, \mathrm{erg \, s^{-1}}$. This is more than one order of magnitude higher than the quiescent luminosity derived from the 2019 \textit{Chandra} observation, $L_\mathrm{X,q} \sim 2 \times 10^{32} \, \mathrm{erg \, s^{-1}}$ (see Sect.~\ref{sec:Chandra_analysis}).

\subsubsection{\textit{NICER}} \label{sec:NICER_spectral_analysis}
We analyzed the \textit{NICER} spectrum in the 1$-$5~keV band, as the background dominates outside this range. The absorption column density was fixed to the same value used in the modeling of the EPIC-pn spectrum (Table~\ref{tab:params_spectrum}). We first described the continuum with an absorbed power-law emission (\texttt{TBabs * pegpwrlw}), resulting in a fit statistic of $\chi^2 = 34.91$ for 45 d.o.f.. The best-fit photon index was $\Gamma = 1.9 \pm 0.2$, and the unabsorbed flux in the 0.5$-$10~keV band was $F_\mathrm{{0.5-10}} = (2.4 \pm 0.2) \times 10^{-12} \, \mathrm{erg \, cm^{-2} \, s^{-1}}$ (corresponding to $L_X \sim 1 \times 10^{34} \, \mathrm{erg \, s^{-1}}$ for $d=6.9 \, \mathrm{kpc}$).
Following the approach used for the \textit{XMM-Newton} spectrum (see Sect.~\ref{sec:XMM_spectral_analysis}), we tested an alternative model replacing the power-law with a Comptonization component (\texttt{TBabs * nthComp}), and fixing the electron temperature at 20~keV (see Sect.~\ref{sec:XMM_spectral_analysis}). The resulting fit yielded a $\chi^2 = 34.90$ for 44 d.o.f., indicating no significant improvement over the power-law model. We also tried the convolution model \texttt{TBabs * (thcomp*bbodyrad)}, but the parameters of the blackbody component were not constrained by the data, likely due to the limited photon statistics. 
The unabsorbed 0.5$-$10~keV flux derived from the power-law model is roughly twice that estimated in the \textit{XMM-Newton} observation performed $\sim$11~days earlier (see Fig.~\ref{Fig:outburst_lc}). This increase may indicate reflare activity, namely luminosity variations commonly observed toward the end of outbursts in several AMSPs \citep[see, e.g.,][and references therein]{Patruno_Watts_2021ASSL}. 

\subsubsection{\textit{Chandra} (2019)} \label{sec:Chandra_analysis}
We fit the \textit{Chandra} spectrum in the 0.3$-$10~keV energy range. Given the low number of counts (see Sect.~\ref{sec:Chandra_obs}), we adopted the Cash statistic (C-statistic; \citealt{Cash_1979ApJ}) for spectral fitting. 
We modeled the spectrum using either an absorbed power law or a blackbody model. All fits included interstellar absorption, with the hydrogen column density fixed to the value reported for the EPIC-pn spectrum (Table~\ref{tab:params_spectrum}).
Using the \texttt{TBabs*pegpwrlw} model, the best-fit photon index was $\Gamma =1.3 \pm 0.4$ -- consistent with the other values reported in both this work, given the large associated uncertainty --, and the unabsorbed 0.5$-$10~keV flux was $F_\mathrm{{0.5-10},q} = 3.8^{+0.8}_{-0.7} \times 10^{-14} \, \mathrm{erg \, cm^{-2} \, s^{-1}}$.  We also fit the spectrum with a blackbody model (\texttt{TBabs*bbodyrad}), including the \texttt{cflux} component to estimate the unabsorbed flux. The best-fit parameters were a blackbody temperature of $\mathrm{kT} =0.9^{+0.2}_{-0.1} \, \mathrm{keV}$, a normalization of $\mathrm{Norm_{bbodyrad}} = 3^{+2}_{-1} \times 10^{-3}$, and an unabsorbed 0.5$-$10~keV flux of $F_\mathrm{{0.5-10},q} = (2.5 \pm 0.5) \times 10^{-14} \, \mathrm{erg \, cm^{-2} \, s^{-1}}$. Simultaneous fits using both a power-law and a thermal component yielded unconstrained parameters due to low photon statistics. However, the small normalization of the blackbody model suggests that the power-law component may be physically preferable. This interpretation is consistent with past works, where the quiescent X-ray spectra of most AMSPs are dominated by a power-law–like component, likely arising from residual accretion or magnetospheric processes related to the NS magnetic field \citep[see, e.g.,][and references therein]{Marino_2022MNRAS}.
In any case, the two flux estimates reported above are consistent with each other within less than 2$\sigma$. Therefore, to conservatively estimate the quiescent luminosity, we adopted the higher flux value obtained from the power-law model. Assuming an upper limit on the source distance of $d \sim 6.9 \, \mathrm{kpc}$ \citep{Altamirano_2010MNRAS}, we derived a quiescent X-ray luminosity of IGR~J17511$-$3057 of $L_\mathrm{X,q} \sim 2 \times 10^{32} \, \mathrm{erg \, s^{-1}}$ in the 0.5$-$10~keV band.

\subsection{Timing analysis} \label{sec:timing_analysis}
\subsubsection{\textit{XMM-Newton}} \label{sec:XMM_timing_analysis}
To correct the photon arrival times for the pulsar's orbital motion in the binary system in the 0.3$-$10~keV \textit{XMM-Newton}/EPIC-pn time series, we adopted the timing solution obtained with \textit{NICER} and \textit{NuSTAR} during this outburst by \citet{Sanna_2025AA}, namely a spin frequency of $\nu=244.83395034(3) \, \mathrm{Hz}$ (corresponding to a spin period of $P_{\mathrm{s}}=0.0040844008709(5) \, \mathrm{s}$), an orbital period of $P_{\mathrm{orb}}=12487.505(2) \, \mathrm{s}$, a projected semi-major axis of $a_1 \sin{i} = 0.275197(8) \, \mathrm{lt}-\mathrm{s}$, and an epoch of passage at the ascending node of $T_\mathrm{asc}=60717.674090(3) \, \mathrm{MJD (TDB)}$ (with respect to the reference epoch $T_\mathrm{ref}=60717.7 \, \mathrm{MJD (TDB)}$). 
The sensitivity to $T_\mathrm{asc}$ that can be obtained by analyzing the \textit{XMM-Newton} observation is $\sim$7~s \citep[e.g.,][]{Caliandro_2012MNRAS}, much lower than the above-reported estimate previously obtained from phase-connected pulsar timing.

\begin{figure}
    \centering
    \includegraphics[width=0.47\textwidth]{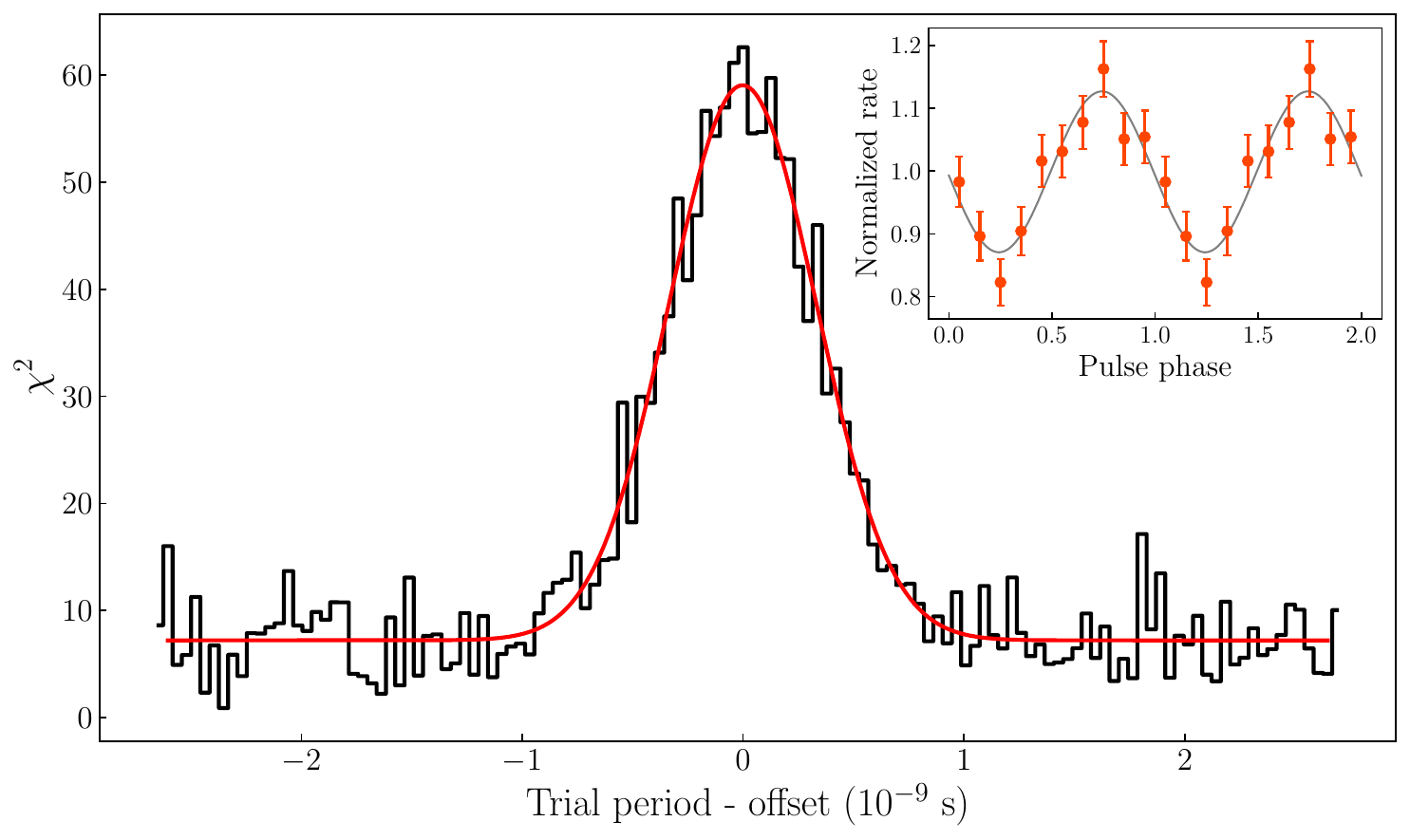}
    \includegraphics[width=0.47\textwidth]{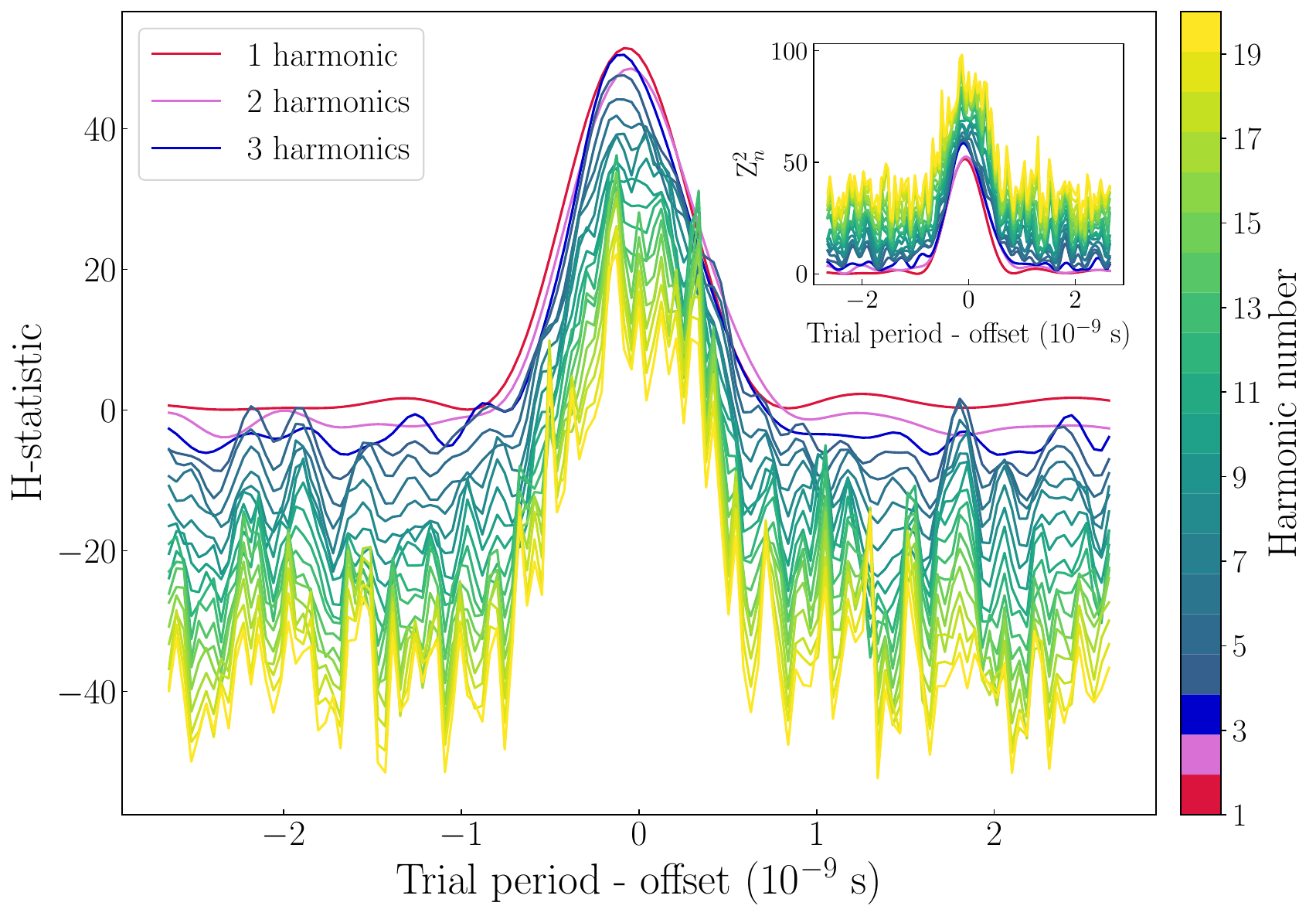}
    \caption{Top panel: Epoch-folding search obtained with 128 trial periods by sampling each period with 10 phase bins. The black points represent the resulting $\chi^2$ values, while the red line shows the best-fit Gaussian model used to determine the spin period. The indicated $x$-axis offset corresponds to $P_{\mathrm{s}}=0.0040844008709 \, \mathrm{s}$. The inset in the top-right corner shows the corresponding pulse profile using 10 phase bins around $P_{\mathrm{s}}$, with $T_\mathrm{ref}=60717.7 \, \mathrm{MJD (TDB)}$ as reference epoch.
    Bottom panel: H-statistic values as a function of the trial periods, computed using the same grid as in the epoch-folding search. For each trial period, we selected the harmonic number that maximized the H-statistic. The red, pink, and dark blue lines represent the values computed for 1, 2, and 3 harmonic terms, respectively. The remaining harmonics, up to 20 following the recommendation of \citet{deJager2010A&A}, are color-coded according to the viridis colormap shown on the right. The inset in the top-right corner shows the corresponding Rayleigh statistics ($Z^2_n$; \citealt{Buccheri1983A&A}) as a function of the trial periods, using the same color scheme.
    See Sect.~\ref{sec:XMM_timing_analysis} for details.} \label{Fig:EFS}
\end{figure}

After correcting the photon arrival times using the values mentioned above, we performed an epoch-folding search (EFS; \citealt{Leahy_1983ApJ, Leahy_1987A&A}) to investigate the presence of periodic signals. We folded the data around the previously mentioned spin frequency. We used $m=10$ phase bins, a period step of $\Delta P = P_\mathrm{s}^2/(2mT) \simeq 4.2 \times 10^{-11} \, \mathrm{s}$ (where $T$ is the observation exposure), and 128 trial periods. 
The resulting EFS, shown in the top panel of Fig.~\ref{Fig:EFS}, reveals a clear peak, from which the best-fit value for the spin period was determined by modeling the $\chi^2$ distribution with a Gaussian function.
We estimated the associated uncertainty in the spin period using the method described by \citet{Leahy_1987A&A}.
The best-fit spin period was $0.0040844009(1) \, \mathrm{s}$ (corresponding to a frequency of $244.833950(6) \, \mathrm{Hz}$), which is consistent within 1$\sigma$ with the earlier estimate. Since we were unable to further improve the precision of this measurement with \textit{XMM-Newton} data, we fixed the spin frequency at $\nu=244.83395034(3) \, \mathrm{Hz}$ for the remaining analysis. Using the maximum $\chi^2$ value of the EFS and the cumulative distribution function of the $\chi^2$ with $m-1$ degrees of freedom, we derived a false alarm probability of $\sim$$4.2 \times 10^{-10}$, corresponding to a single-trial statistical significance exceeding $6$$\sigma$. This clearly indicates the presence of X-ray pulsations, even though the source luminosity had already decreased by more than two orders of magnitude compared to previous \textit{INTEGRAL}, \textit{NICER}, and \textit{NuSTAR} observations (see Fig.~\ref{Fig:outburst_lc}).

Up to five harmonic components have been reported to significantly contribute to the pulse profiles of IGR~J17511$-$3057 (\citealt{Papitto_2010MNRAS}; see also \citealt{Riggio_2011A&A, Papitto_2016A&A}; \citealt{Sanna_2025AA}). To determine the optimal number of harmonic terms needed to describe the X-ray pulsations detected during the \textit{XMM-Newton} observation, we computed the H-statistic \citep[e.g.,][and references therein]{deJager2010A&A} as a function of the trial periods.
The results, shown in the bottom panel of Fig.~\ref{Fig:EFS}, indicate that the signal is best described by a single harmonic component, consistent with previous findings that the first harmonic dominates the pulse profile \citep{Papitto_2010MNRAS, Riggio_2011A&A, Papitto_2016A&A, Sanna_2025AA}. The harmonic content is likely determined by geometric effects and the observer viewing angle \citep[e.g.,][]{Poutanen_Beloborodov_2006MNRAS, Das_2025ApJ}.

Due to the low observed X-ray flux, it was not possible to use the \textit{XMM-Newton} data to derive an independent coherent timing solution. Consequently, we were unable to study the evolution of the X-ray pulsation phase. Instead, we folded the entire observation across different energy ranges using 10 phase bins around $P_{\mathrm{s}}$ and using the corresponding reference epoch $T_\mathrm{ref}$, as defined above (see the inset in the top panel of Fig.~\ref{Fig:EFS}).
We modeled the folded pulse profiles with one harmonic term, as suggested by the H-test (see bottom panel of Fig.~\ref{Fig:EFS}).
Additionally, we performed an energy-resolved analysis of the 0.3–10~keV pulsations to explore their spectral dependence. The results are summarized in Table~\ref{tab:energy_ranges}. We found that X-ray pulsations are significantly stronger at lower energies. In the 0.3$-$3~keV and 0.3$-$5~keV bands, pulsations are detected at $>5$$\sigma$ significance. In contrast, the 3–10~keV and 5–10~keV bands show much weaker or absent pulsations, with significances of $2.1$$\sigma$ and $0.3$$\sigma$, respectively. This may be due either to lower photon statistics at higher energies, as indicated by the source and background counts in Table~\ref{tab:energy_ranges}, or to an intrinsically lower pulsed amplitude in these bands. In both cases, the results indicate that the pulsed signal is dominated by soft X-ray photons.

\begin{table}
\renewcommand{\arraystretch}{1.2}
\centering
\caption{Properties of X-ray pulsations with energy.} \label{tab:energy_ranges}
\begin{tabular}{cccccc} 
\hline
Band & $N_\mathrm{src}$ & $N_\mathrm{bkg}$ & A (\%) & p-value & S ($\sigma$)\\
\hline
0.3$-$10~keV & 6003 & 3668 & $32.9 \pm 4.8$ & $1.5 \times 10^{-8}$ & $5.7$ \\
\hline
0.3$-$3~keV  & 4269 & 2838 & $42.0 \pm 6.7$& $1.4 \times 10^{-7}$ & $5.3$ \\
3$-$10~keV   & 1733 & 829 & $21.8 \pm 6.5$& $3.2 \times 10^{-2}$ & $2.1$ \\
\hline
0.3$-$5~keV  & 5194 & 3154 & $37.1 \pm 5.1$& $2.1 \times 10^{-9}$ & $6.0$ \\
5$-$10~keV   & 809 & 514 & $25 \pm 14$& $7.6 \times 10^{-1}$ & $0.3$ \\
\hline
\end{tabular}
\tablecomments{$A$ is the background-subtracted amplitude. $N_\mathrm{src}$ and $N_\mathrm{bkg}$ are the source and background counts, both extracted from a 11-pixel-wide strip (see Sect.~\ref{sec:XMM_data}). $A$ was obtained by scaling the observed amplitude by a factor $N_\mathrm{src}/(N_\mathrm{src} - N_\mathrm{bkg})$. The p-value is the false alarm probability, and S indicates the corresponding single-trial statistical significance in units of $\sigma$.}
\end{table}

\subsubsection{\textit{NICER}} \label{sec:NICER_timing_analysis}
After correcting the photon arrival times for the NS orbital motion using the orbital parameters provided in Sect.~\ref{sec:XMM_timing_analysis}, we searched for a coherent signal in the Leahy-normalized Fourier power spectrum \citep{Leahy_1983ApJ} of the 0.3$-$10~keV \textit{NICER} data. The frequency resolution of the power spectrum computed over the full observation duration ($T_{\mathrm{obs}}\sim6.5$~ks, $\delta\nu=1.5\times10^{-4}$~Hz) is much coarser than the accuracy in the available pulsar ephemeris. Consequently, the search for a coherent signal at the known NS spin period was constrained to a single-trial analysis. For such a search, the power threshold corresponding to a 3$\sigma$ confidence level (false alarm probability $p = 2.7 \times 10^{-3}$)\footnote{The false alarm probability $p$ is defined as $1 - C$, where $C$ is the confidence level \citep[e.g.,][]{Vaughan_1994}.} is approximately $2 \ln{(1/p)} \simeq 11.8$. Given a total of $N_\mathrm{ph} = 2172$ counts and a Nyquist frequency of 1~kHz, the observed power of $\sim$6.5 translates into a 3$\sigma$ upper limit on the pulsed amplitude of $\sim$18\%, following the approach of \citet{Vaughan_1994}. Accounting for the average background rate ($R_{\mathrm{bkg}} \sim 0.16$~cts~s$^{-1}$), this upper limit increases to $\sim$31\% in the 0.3$-$10~keV band.
Although the background-subtracted pulsed amplitude estimated with \textit{XMM-Newton} in the same energy band slightly exceeds this limit, it remains consistent within the uncertainties (see Table~\ref{tab:energy_ranges}). Moreover, \citet{Sanna_2025AA} reported a pulsed amplitude of $\sim$19\% with \textit{NICER} at the onset of the outburst. Therefore, the non-detection of coherent pulsations in the \textit{NICER} observation from March 15, 2025 cannot be taken as definitive evidence of a significant decrease in the pulse amplitude. However, this could result from limited sensitivity, since the substantially lower average count rate of $\sim$1~cts~s$^{-1}$, in contrast to the rate of 52~cts~s$^{-1}$ recorded early in the outburst, suggests a substantial decline in the mass accretion rate onto the NS during the later stages of the outburst.

\subsection{Light curve}
\begin{figure*}
    \centering
    \includegraphics[width=0.9\textwidth]{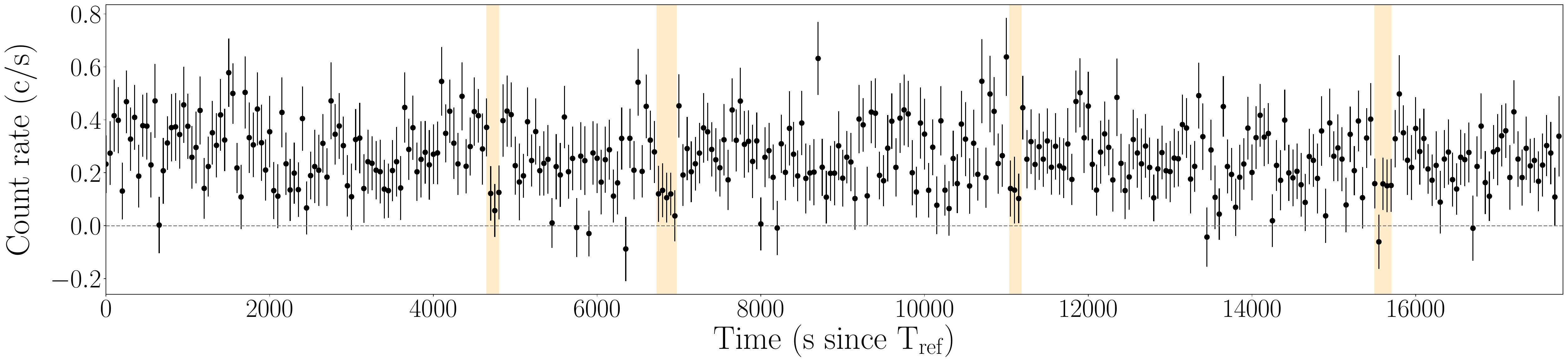}
    \caption{Background-subtracted light curve from the three EPIC instruments over the time interval of simultaneous coverage, binned at a 50-s resolution. The yellow-shaded regions highlight potential low modes (see Sect.~\ref{sec:intro} for details). Time is expressed in seconds relative to the reference time $\mathrm{T_{ref}}=$60738.0582657791310339~MJD (TDB).} \label{Fig:EPIC_lc}
\end{figure*}

\begin{figure}
    \centering
    \includegraphics[width=0.47\textwidth]{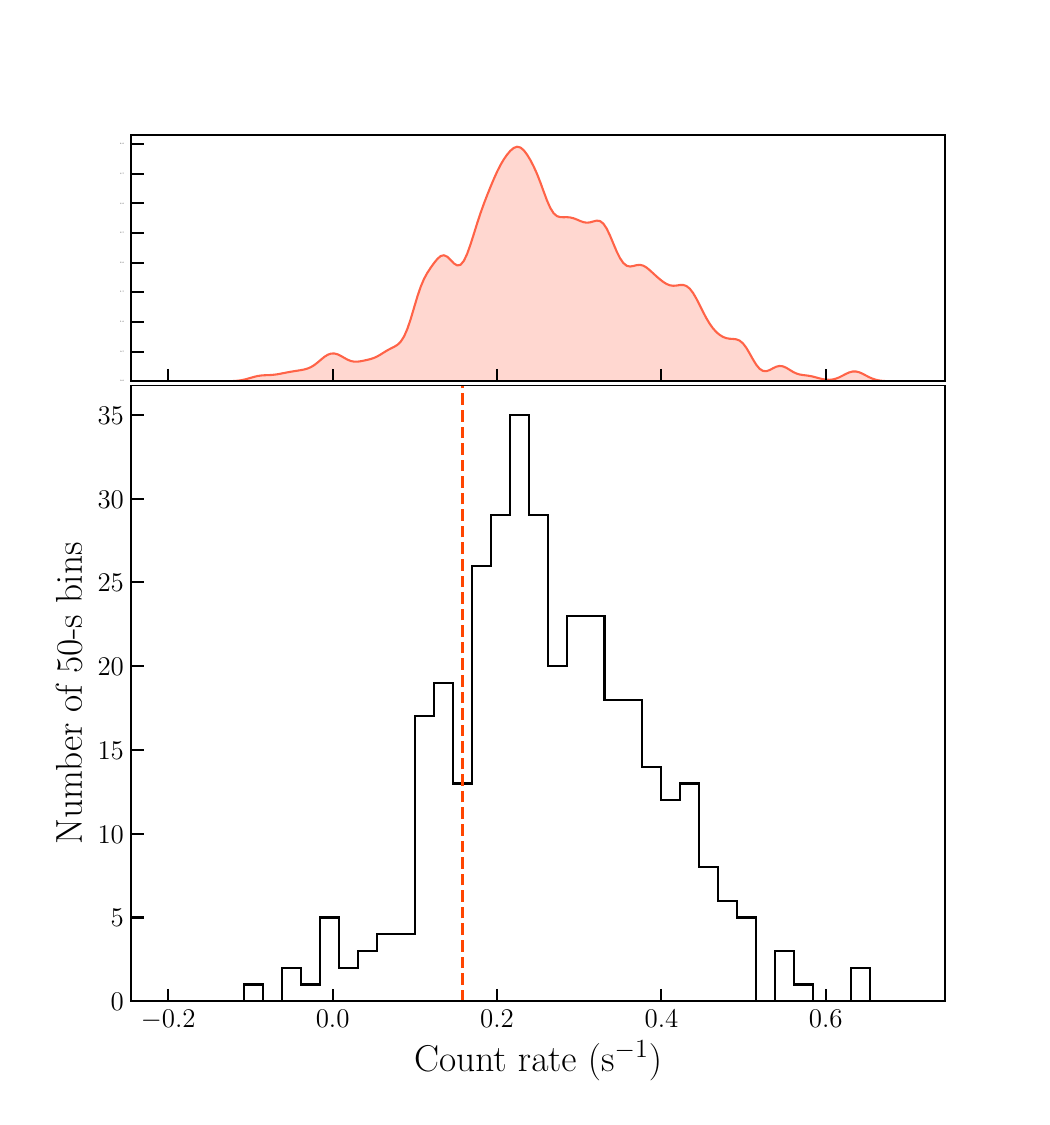}
    \caption{Distribution of count rates obtained from the background-subtracted EPIC light curve (Fig.~\ref{Fig:EPIC_lc}), binned with a time resolution of 50 s. A potential ``transition'' threshold between high and low modes is indicated by a dashed orange line. Negative count rates can occur in bins where the source signal is very low or absent and the background dominates, as a result of the background subtraction. The upper panels display the Kernel Density Estimation curve, providing a smoothed representation of the count rate distribution.} \label{Fig:EPIC_counts_distribution}
\end{figure}

Figure~\ref{Fig:EPIC_lc} shows the background-subtracted light curve from the three EPIC instruments over the time interval of simultaneous coverage, binned at a 50-s resolution. The light curves from each detector were extracted using the \texttt{epiclccorr} task and combined with the \texttt{lcmath} task. The distribution of count rates is shown in Fig.~\ref{Fig:EPIC_counts_distribution}.
Our findings are discussed in Sect.~\ref{sec:tMSPs}.

\section{Discussion} \label{sec:discussion}
In this paper, we reported on the detection of coherent X-ray pulsations from the AMSP IGR~J17511$-$3057 observed with \textit{XMM-Newton} during a ToO observation conducted about twenty days after the onset of its 2025 outburst (see Fig.~\ref{Fig:outburst_lc}). Given that past outbursts from this source typically lasted for $\sim$25$-$30~days \citep[e.g.,][]{Papitto_2010MNRAS, Riggio_2011A&A, Papitto_2016A&A}, our observation likely captured the system during its final stages of activity.
Spectral analysis revealed a 0.5$-$10~keV unabsorbed X-ray flux of $F_X = (1.23 \pm 0.04) \times 10^{-12} \, \mathrm{erg \, cm^{-2} \, s^{-1}}$. Assuming the upper limit on the source distance of $d \sim 6.9 \, \mathrm{kpc}$ \citep{Altamirano_2010MNRAS}, this corresponds to an X-ray luminosity of $L_X \sim 7 \times 10^{33} \, \mathrm{erg \, s^{-1}}$. To put this measurement into context, we analyzed an archival \textit{Chandra} observation performed in 2019 during quiescence, from which we derived a 0.5$-$10~keV luminosity of $L_{\mathrm{X,q}} \sim 2 \times 10^{32} \, \mathrm{erg \, s^{-1}}$ (see Sect.~\ref{sec:Chandra_analysis}).
This indicates that the source was still over an order of magnitude brighter than its quiescent level at the time of the \textit{XMM-Newton} observation. In the following, we discuss our results within the framework of the standard accretion–propeller boundary and examine the increasingly subtle distinction between AMSPs during the late phases of the X-ray outbursts and tMSPs in their sub-luminous disk state.

\subsection{Pushing beyond the propeller limit} \label{sec:propeller}
The detection of coherent millisecond pulsations from AMSPs is thought to provide strong evidence for ongoing accretion onto the NS magnetic poles. Specifically, accretion-powered X-ray pulsations imply that the inner disk radius must be truncated not significantly beyond the co-rotation radius, defined as the distance at which the angular velocity of a disk matter element in a Keplerian orbit matches that of the co-rotating system formed by the NS and its magnetosphere. The co-rotation radius is given by
\begin{equation}
    \mathrm{R_{co}} = \left[ \frac{G M_{\mathrm{ns}}}{(2 \pi \nu)^2} \right]^{1/3} \simeq 1.68 \times 10^{6} \, \mathrm{M_{1.4}} \, \mathrm{P_{ms}}^{2/3} \, \mathrm{cm},
\end{equation}
where $\mathrm{M_{1.4}}$ is the NS mass ($M_{\mathrm{ns}}$) in units of 1.4 $\mathrm{M_\odot}$ and $\mathrm{P_{ms}}$ is the spin period in ms. For a 1.4 $\mathrm{M_\odot}$ NS and a spin period of $\sim$4.1 ms, $\mathrm{R_{co}} \simeq 43$~km.
If the inner disk were truncated beyond this radius, the rapidly rotating magnetosphere would act as a centrifugal barrier, preventing accretion in the so-called propeller regime \citep{Illarionov_Sunyaev_1975A&A}.
To estimate the mass accretion rate during the \textit{XMM-Newton} observation, we extrapolated the unabsorbed X-ray flux in the 0.1-100~keV band, $F_\mathrm{X} = (1.44 \pm 0.06) \times 10^{-12} \, \mathrm{erg \, cm^{-2} \, s^{-1}}$. For $d=6.9 \, \mathrm{kpc}$ and assuming isotropic emission, the accretion rate is $\dot{M} = 4 \pi d^2 F_X R_{\mathrm{ns}} /(G M_{\mathrm{ns}}) \simeq 4 \times 10^{13} \, \mathrm{g \, s^{-1}}$ for a 1.4 M$_\odot$ NS with a $R_{\mathrm{ns}}=10$~km radius.
The truncation radius of the disk, $R_{\mathrm{m}}$, is assumed to be $\xi$ times the Alfvén radius \citep[e.g.,][]{Ghosh_Lamb_1979ApJ} and is expressed as
\begin{equation}
    R_{\mathrm{m}} = \xi  \left[ \frac{(B_{\mathrm{s}} R_{\mathrm{ns}}^3)^4}{(2 G M_{\mathrm{ns}} \dot{M}^2)} \right]^{1/7},
\end{equation}
where $B_{\mathrm{s}}$ is the NS magnetic field strength at surface, and $G$ is the gravitational constant. Requiring $R_{\mathrm{m}} < R_{\mathrm{co}}$ for accretion to proceed, we derived an upper limit on the magnetic field:
\begin{equation}
    B_{\mathrm{s}} < \xi^{-7/4} R_{\mathrm{co}}^{7/4} (2\,G\,M\,\dot{M}^2)^{1/4} R_{\mathrm{ns}}^{-3} \sim 4 \times 10^{7} \, \mathrm{G},
\end{equation}
where we adopted $\xi=0.5$ \citep[see, e.g.,][]{Campana_2018A&A}.
However, \citet{Sanna_2025AA} reported constraints on the magnetic field strength of $7.6 \times 10^{8} \, \mathrm{G}  < B_{\mathrm{s}} < 1.1 \times 10^{9} \, \mathrm{G}$ based on the secular spin-down of the pulsar during quiescence, interpreted as due to magnetic dipole radiation. This field strength exceeds the value required to maintain $R_{\mathrm{m}} < R_{\mathrm{co}}$ during the \textit{XMM-Newton} observation. We note that both field strength estimates rely on simplifying assumptions. Our limit depends on the adopted conversion from the X-ray luminosity to $\dot{M}$, on the assumed NS parameters, and on the value of $\xi$ relating the magnetospheric and Alfvén radii, while the secular spin-down estimate assumes a specific torque mechanism in quiescence, a canonical moment of inertia, and does not account for possible spin-up episodes during outbursts. Although it is difficult to quantitatively assess the uncertainties associated with these assumptions, the two estimates for the magnetic field intensity still differ by about one to two orders of magnitude. If accretion is still ongoing — as supported by the detection of coherent X-ray pulsations — and the average long-term spin down is a good tracer of the NS magnetic field, then our findings would suggest that the truncation radius of the disk can remain within the co-rotation radius even at very low X-ray luminosities, comparable to those seen in tMSPs during their sub-luminous disk state (see Sections~\ref{sec:intro} and \ref{sec:tMSPs}). This result is consistent with what was found by \citet{Patruno_2016}, whose multi-wavelength analysis of the reflaring phases of the prototype of AMSP SAX~J1808.4$-$3658 showed that the persistence of pulsations and a stable hard X-ray spectrum imply a disk truncated near the co-rotation radius during the late stages of the outburst.

Different mechanisms have been proposed to explain how accretion can persist in such a low-luminosity regime \citep[e.g.,][]{Archibald_2015, Papitto_2015MNRAS, Bult_2019ApJ}. One possibility is that, at low accretion rates, the inner accretion flow becomes radiatively inefficient, so only a fraction of the accretion power is emitted in X-rays while most of it is instead lost through outflows \citep[e.g.,][]{Zuo_Cugno_2025ApJ}. As a result, the actual mass accretion rate -- and thus the ram pressure acting on the NS magnetosphere -- may be greater than inferred from the observed X-ray luminosity, allowing the disk to remain close to the co-rotation radius.
Another possibility is the ``trapped disk'' configurations \citep{Spruit_Taam_1993ApJ, DAngelo_2010MNRAS, DAngelo_2012MNRAS}, in which the magnetosphere rotates only slightly faster than the inner disk. In this case, the centrifugal barrier is not sufficient to expel the inflowing matter, allowing material to accumulate near the magnetospheric radius and accrete episodically.
On the other hand, three-dimensional magnetohydrodynamic simulations have revealed that accretion and ejection can coexist in the propeller regime, even when the inner disk lies beyond the corotation radius \citep[see][and references therein]{Romanova_2018NewA}.

During the 2025 outburst, \citet{Sanna_2025AA} reported average background-subtracted amplitude of $\sim$19\% for the first harmonic with \textit{NICER} (0.3$-$10~keV) and \textit{NuSTAR} (3-40~keV). Specifically, the \textit{NICER} data showed a rise from $\sim$16\% at the low energies up to $\sim$21\%, while the \textit{NuSTAR} pulse profiles revealed a peak amplitude of $\sim$22\% in the 3$-$6~keV range, decreasing at the higher energies. The results are broadly consistent with previous outbursts \citep{Papitto_2010MNRAS, Riggio_2011A&A, Papitto_2016A&A}. In terms of energy dependence, \textit{XMM-Newton} and \textit{RXTE} observations during the 2009 outburst showed that the amplitude of the first harmonic increased with energy up to $\sim$20–25\% in the 3$-$6~keV range, and then gradually decreased at higher energies \citep{Papitto_2010MNRAS}.
The 2025 \textit{XMM-Newton} observation presented in this work revealed a background-subtracted amplitude of $\sim$33\% in the 0.3$-$10~keV band, peaking at $\sim$42\% in the softer 0.3$-$3~keV band. Pulsations were not significantly detected above this energy (see Table~\ref{tab:energy_ranges}). Such high pulse amplitudes are rare among AMSPs, although similar values have been observed in a few systems -- such as  XTE~J1807$-$294 (up to 27\% in the 2.5$-$30~keV band; \citealt{Patruno_2010ApJ}), Swift~J1749.4$-$2807 (up to 23\% in the 2$-$10~keV band; see, e.g., \citealt{Altamirano_2011ApJ, Sanna_2022}), and IGR~J17379$-$3747 (up to 70\% in the 0.4$-$6~keV band; \citealt{Sanna_2018AA_IGRJ17379-3747, Bult_2019ApJ}). Interestingly, the latter also displayed pulsations down to luminosities as low as $\sim$$5 \times 10^{33} \, \mathrm{erg \, s^{-1}}$ \citep{Bult_2019ApJ}, similar to the results presented in this paper for IGR~J17511$-$3057. According to \citet{Poutanen_Beloborodov_2006MNRAS}, such large pulse amplitudes can be explained by a favorable viewing geometry, in which the sum of the inclination angle and spot colatitude is close to $90^\circ$.
This geometric configuration may explain the unusually high pulse amplitude observed in these two AMSPs, which likely enables the detection of X-ray pulsations even at low accretion rates, as further discussed in the following paragraph.

The low-luminosity X-ray pulsations observed from IGR~J17511$-$3057 in this work and IGR~J17379$-$3747 by \citet{Bult_2019ApJ} are rare among AMSPs. This rarity could be partly attributed to the sensitivities of the X-ray telescopes, as well as the relatively low pulse amplitudes of most systems. A useful comparison can be made with the extensively studied AMSP SAX~J1808.4$-$3658 \citep{Wijnands_VanDerKlis_1998Natur}, which has exhibited ten $\sim$1-month-long outbursts every $\sim$$3$ years since its discovery as the first system of this kind \citep[see][and references therein]{Illiano_2023ApJ}. The amplitude of X-ray pulsations observed from SAX~J1808.4$-$3658 typically ranges between $\sim$3-7\% \citep[e.g.,][]{Hartman_2008ApJ, Hartman_2009, Patruno_2012, Sanna_2017, Bult_2020ApJ, Illiano_2023ApJ}, although larger values have been observed in the late stages of the outburst \citep[e.g.,][]{Hartman_2008ApJ, Patruno_2009ApJ_SAXJ1808, Bult_2020ApJ, Illiano_2023ApJ, Ballocco_in_prep}. Although X-ray pulsations from SAX~J1808.4$-$3658 have been consistently detected across its outbursts, including during the late accretion phases known as reflares, they have only been observed down to luminosities of $\sim$$(2-6) \times 10^{34} \, \mathrm{erg \, s^{-1}}$ in the 0.5$-$10~keV band \citep[e.g.,][]{Hartman_2008ApJ, Patruno_2009ApJ_SAXJ1808, Bult_2019ATel13001, Ballocco_in_prep}, and never at luminosities as low as those seen in IGR~J17511$-$3057 and IGR~J17379$-$3747.
Considering the single-trial sensitivity required for a 3$\sigma$ confidence level detection \citep[see][]{VanDerKlis_1988, Vaughan_1994}, a typical pulse amplitude of 5\% in SAX~J1808.4$-$3658 would, in principle, be detectable down to a minimum 0.5$-$10~keV luminosity of $\sim$$3 \times 10^{33} \, \mathrm{erg \, s^{-1}}$ in a 50~ks \textit{XMM-Newton} observation (adopting the commonly used distance of 3.5~kpc from \citet{Galloway_2006}, although \citet{Galloway_2024MNRAS} recently suggested a revised distance of $2.7$~kpc). 
However, in practice, capturing SAX~J1808.4$-$3658 at such low luminosity before it fades into quiescence remains observationally challenging, especially due to the presence of the typical reflaring phase. This underscores how AMSPs with intrinsically higher pulse amplitudes, such as IGR~J17511$-$3057 and IGR~J17379$-$3747, represent a unique opportunity for studying X-ray pulsations at the lowest accretion luminosities during the late stages of their outbursts. 

Similar to IGR~J17379$-$3747 \citep{Bult_2019ApJ}, the X-ray spectrum of IGR~J17511$-$3057 becomes significantly softer in the late stages of the 2025 outburst. During the \textit{XMM-Newton} observation, we found a photon index of $\Gamma \simeq 2.2$ (Table~\ref{tab:params_spectrum}), in contrast to the harder spectra reported earlier in the outburst with \textit{NICER} and \textit{NuSTAR} ($\Gamma \simeq 1.7-1.8$; \citealt{Sanna_2025AA}). This spectral softening toward the end of the outburst mirrors trends seen in both black hole \citep[e.g.,][]{Wu_2008ApJ, Plotkin_2013ApJ} and NS binaries \citep[see][and references therein]{Wijnands_2015MNRAS}. For NS binaries, possible explanations for this behavior involve residual accretion onto the NS surface \citep[e.g.,][]{Campana_1998A&ARv} or the gradual cooling of the stellar crust, previously heated during the outburst \citep[e.g.,][]{Brown_1998ApJ}. However, \citet{Allen_2015ApJ} found that the spectral softening in the NS LMXB SAX~J1750.8$-$2900 is best explained by ongoing low-level accretion, based on the presence of rapid variability and correlated spectral changes, which are inconsistent with the timescales expected from crustal cooling alone.
The detection of coherent X-ray pulsations in both IGR~J17511$-$3057 (this work) and IGR~J17379$-$3747 \citep{Bult_2019ApJ} at low-luminosity levels supports the presence of a residual accretion.

The findings discussed in this section support the idea that at least some AMSPs can sustain accretion-powered X-ray pulsations down to remarkably low luminosities, challenging the traditional boundary between the accretion and propeller regimes. 

\subsection{Comparison with transitional millisecond pulsars} \label{sec:tMSPs}
Despite a two-order-of-magnitude decline compared to earlier outburst levels (Fig.~\ref{Fig:outburst_lc}), the 2025 \textit{XMM-Newton} luminosity remained significantly above the quiescent level and falls within the typical range observed in the intermediate sub-luminous disk state of tMSPs. Motivated by this, we searched for potential bimodality between high and low intensity levels in the X-ray light curve. Figure~\ref{Fig:EPIC_lc} shows the background-subtracted light curve from the three EPIC instruments, binned at a 50-s resolution. Intervals that visually resemble low-mode episodes are highlighted in yellow. However, the lack of a clear bimodal distribution in the count rates (bottom panel of Fig.~\ref{Fig:EPIC_counts_distribution}) prevents us from drawing firm conclusions.
The estimated upper limit on the distance of IGR~J17511$-$3057 is 6.9~kpc, notably larger than those of confirmed tMSPs that have exhibited prolonged sub-luminous disk states, namely PSR~J1023$+$0038 and XSS~J12270$-$4859 (both at $\sim$1.4~kpc; \citealt{Deller2012, deMartino_2020MNRAS}). Most tMSP candidates with clear bimodality are located within $\sim$4~kpc \citep{CotiZelati_2019A&A, CotiZelati2024, Bogdanov_2015ApJ, Bogdanov_2016ApJ, Gusinskaia_2025MNRAS, Illiano_2025A&A}, whereas more distant systems, such as 4FGL~J0407.7$-$5702 (6.9-12.5~kpc; \citealt{Miller_2020}) and 4FGL~J0639.1$-$8009 (6.6-13.3 kpc; \citealt{Kyer_2025ApJ}), display weaker or absent low-mode activity despite other tMSP-like characteristics.
A similar large distance ($\sim$8~kpc) is estimated for the AMSP IGR~J17379$-$3747, for which low-luminosity X-ray pulsations have been observed \citep{Bult_2019ApJ}. This supports the possibility that short-lived sub-luminous disk states may remain undetected in more distant systems, as larger distances reduce the signal-to-noise ratio. In addition, intrinsic factors such as system geometry and mass transfer rate may affect the strength of mode switches and therefore their detectability.

The average X-ray spectrum of tMSPs in the sub-luminous disk state is usually well described by an absorbed power-law with a photon index of $\Gamma \sim 1.4-1.7$ \citep[see][and references therein]{Papitto_deMartino_2022ASSL}, occasionally softening to $\Gamma \simeq 2$ during low modes \citep{Campana_2016A&A, CotiZelati2018}. For IGR~J17511$-$3057, the photon index at the onset of the 2025 outburst was $\Gamma \simeq 1.7-1.8$, while the subsequent low-luminosity \textit{XMM-Newton}/EPIC spectrum was significantly softer ($\Gamma \simeq 2.2$; see also Sect.~\ref{sec:propeller}). 
Although this value is higher than typically observed in tMSPs in the sub-luminous disk state, a similar spectral softening was reported for the tMSP IGR~J18245$-$2452 \citep{Linares_2013MNRAS} and for the AMSP IGR~J17379$-$3747 \citep{Bult_2019ApJ}.

During the 2009 outburst, IGR~J17511$-$3057 showed a clear energy dependence of the pulsed signal (see Sect.~\ref{sec:propeller}), with the first harmonic amplitude increasing with energy up to $\sim$3-6~keV \citep{Papitto_2010MNRAS}.
A similar behavior was reported during the 2025 outburst with \textit{NICER} and \textit{NuSTAR} \citep{Sanna_2025AA}.
However, our \textit{XMM-Newton} observation during the subsequent low-luminosity phase in 2025 shows that pulsations were significantly detected only at low energies, with the highest amplitude in the 0.3$-$3~keV range, and no detection above $3\sigma$ at higher energies (Table~\ref{tab:energy_ranges}). 
This soft energy dependence differs from the trend observed in the tMSP PSR~J1023$+$0038 during the sub-luminous disk state, where the pulsed amplitude increases with energy from optical to X-rays \citep{Papitto_2019ApJ, Miraval_Zanon_2022, Baglio_2025ApJ}. 

Despite this discrepancy, the overall phenomenology of IGR~J17511$-$3057 in its faint accretion state -- characterized by coherent X-ray pulsations, low X-ray luminosity, and spectral softening -- shares similarities with the sub-luminous disk state of tMSPs. Overall, the results presented in this paper not only challenge the traditional boundary between the accretion and propeller regimes, but also blur the distinction between the late stages of X-ray outbursts in some AMSPs and the intermediate sub-luminous disk state of tMSPs.

\subsection{Radio observations} \label{sec:discussion_radio}
Motivated by the possibility that IGR~J17511$-$3057 might produce radio continuum emission associated with residual accretion or weak outflows, we conducted radio observation of IGR~J17511$-$3057 with ATCA on 2025 April 12. This followed a drop in accretion activity, as indicated by \textit{NICER} data from 2025 March 15, which showed an average count rate of $\sim$1~cts~s$^{-1}$ -- significantly lower than the 52~cts~s$^{-1}$ observed at the outburst peak (see Sect.~\ref{sec:NICER}) -- and revealed no significant X-ray pulsations, although this may also be due to the limited sensitivity (see Sect.~\ref{sec:NICER_timing_analysis}). However, \textit{Chandra} data during quiescence in 2019 revealed a very low X-ray luminosity ($L_\mathrm{X,q} \sim 2 \times 10^{32} \, \mathrm{erg \, s^{-1}}$ in the 0.5$-$10~keV band), over an order of magnitude below that during the \textit{XMM-Newton} and \textit{NICER} observations, suggesting that the source may still have been in a faint accretion state during ATCA observations. We did not identify any radio counterpart, with a 3$\sigma$ upper limit of 60\,$\mu$Jy\,beam$^{-1}$ at 5.5~GHz on the total source flux density. 

Radio continuum emission has been detected in only a few AMSPs during their X-ray outburst, where it is interpreted as synchrotron radiation from compact, partially self-absorbed jets \citep[e.g.,][and references therein]{Gaensler_1999ApJ, Tudor_2017MNRAS}. The first detection was obtained from the first discovered and best-studied AMSP SAX~J1808.4$-$3658, with flux densities of $\sim$0.8~mJy at 4.8~GHz during its 1998 outburst with no further detection at later epochs, showing a variable behavior \citep{Gaensler_1999ApJ}. Similar detections were reported for XTE~J0929$-$314 at $\sim$0.3~mJy at 4.86~GHz \citep{Rupen_2002IAUC} and for IGR~J00291$+$5934,  which displayed variable radio emission during its 2004 outburst, with flux densities ranging from $\sim$1.1~mJy (15~GHz) to $\sim$0.17–0.25~mJy (4.86–5~GHz), while it was not detected during its 2008 outburst, with upper limits of $\sim$0.16~mJy at 5~GHz \citep[e.g.,][and references therein]{Lewis_2010A&A}. A bright transient radio counterpart was also detected for the tMSP IGR~J18245$-$2452 during its 2013 accretion outburst, with flux densities of $\sim$0.6~mJy and $\sim$0.8~mJy at 5.5 and 9~GHz, respectively \citep{Pavan_2013ATel}. In contrast, several AMSPs such as HETE~J1900.1$-$2455, Swift~J1756.9$-$2508, yielded no radio detections despite searches during their outburst \citep[see][and references therein]{Patruno_Watts_2021ASSL}.

In light of the comparison between IGR~J17511$-$3057 and the tMSPs in the sub-luminous disk state (Sect.~\ref{sec:tMSPs}), we note that the radio emission in the latter sources is attributed to synchrotron-emitting outflows. The prototype of tMSP in the sub-luminous disk state, PSR~J1023$+$0038, shows a highly variable, flat-spectrum radio emission \citep{Deller_2015, Bogdanov_2015ApJ_J1023}, with an average flux density of $\sim$56~$\mu$Jy at 10~GHz during the X-ray high modes, and short-lived flares reaching up to an average radio flux density $\sim$235~$\mu$Jy at 10~GHz during the low modes (\citealt{Bogdanov2018}; see also \citealt{Baglio_CotiZelati_2023A&A}). 
Hints for a similar anti-correlated radio/X-ray variability pattern were recently found for the promising candidate tMSP 3FGL~J1544.6$-$1125 \citep{Gusinskaia_2025MNRAS}, along with 
remarkably variable radio flux densities ranging from $< 13.8$ (at 3$\sigma$ confidence level) to $\sim$48~$\mu$Jy in the X-band (8$-$12~GHz) \citep{Jaodand_2021ApJ}, and from $\sim$29 to 57~$\mu$Jy in the C-band (4$-$8~GHz) during X-ray low modes \citep{Gusinskaia_2025MNRAS}, with a $3\sigma$ upper limit of $\sim$8~$\mu$Jy in the same band at a later epoch despite a nearly constant X-ray flux \citep{Illiano_2025A&A}.

Given the estimated source's distance, the 3$\sigma$ upper limit of 60~$\mu$Jy\,beam$^{-1}$ at 5.5~GHz obtained for IGR~J17511$-$3057 is not stringent enough to rule out the presence of a radio counterpart when compared with the radio flux densities observed from AMSPs during outburst or from tMSPs in the sub-luminous disk state. Deeper and more sensitive observations will therefore be needed to probe weaker or variable radio emission and to search for possible coherent pulsations. Interferometric radio observations, such as those performed with ATCA, cannot indeed resolve the individual pulses typical of most radio MSPs, which are characterized by low duty cycles and narrow pulse widths \citep[e.g.,][]{Manchester_2005AJ_ATNF}. Future time-resolved radio monitoring will be crucial to distinguish between a quiescent AMSP potentially hosting an active radio pulsar -- as observed for the tMSP IGR~J18245$-$2452 \citep{Papitto_2013Natur}, and suggested by indirect evidence such as the long-term spin-down observed in many AMSPs \citep[e.g.,][and references therein]{Illiano_2023ApJ} -- and a tMSP-like sub-luminous disk state in IGR~J17511$-$3057.

\section{Conclusions} \label{sec:conclusions}
In this paper, we report on \textit{XMM-Newton} and \textit{NICER} observations of the AMSP IGR~J17511$-$3057 during the late stage of its 2025 outburst, complemented by archival 2019 \textit{Chandra} data during quiescence and recent ATCA radio observations. Our main findings can be summarized as follows:
\begin{itemize}
    \item Low-luminosity pulsations: coherent X-ray pulsations were detected with \textit{XMM-Newton} at an unabsorbed flux of $\sim$$1.2 \times 10^{-12} \, \mathrm{erg \, cm^{-2} \, s^{-1}}$ (0.5$-$10~keV), implying an X-ray luminosity of $\sim$$7 \times 10^{33} \, \mathrm{erg \, s^{-1}}$ (assuming a distance equal to the estimated upper limit of $6.9$~kpc). At this luminosity, the source is expected to be in the propeller regime, yet pulsations persist. No significant pulsations were detected in subsequent \textit{NICER} observations, likely due to sensitivity limits and/or decreased accretion activity.
    \item High pulse amplitude: the X-ray pulsations reached a peak amplitude of $\sim$42\% in the 0.3$-$3~keV band. Similar high pulse amplitudes were reported for the AMSP IGR~J17379$-$3747, which also exhibited low-luminosity X-ray pulsations \citep{Bult_2019ApJ}. The high pulse amplitude is likely due to a favorable geometric configuration, enabling the detection of X-ray pulsations even beyond the traditional accretion–propeller boundary.
    \item Spectral softening: Compared to earlier outburst phases \citep{Sanna_2025}, the \textit{XMM-Newton} spectrum showed significant spectral softening, consistent with trends seen in both NS and black hole binaries toward the end of the outburst. Similar behavior was also observed at the end of the 2013 outburst of the tMSP IGR~J18245$-$2452 \citep{Linares_2013MNRAS}.
    \item X-ray quiescent luminosity: The 2019 \textit{Chandra} spectrum, well described by an absorbed power-law model, revealed a quiescent unabsorbed flux of $\sim$$4 \times 10^{-14} \, \mathrm{erg \, cm^{-2} \, s^{-1}}$ (0.5$-$10~keV), corresponding to a quiescent X-ray luminosity of $\sim$$2 \times 10^{32} \, \mathrm{erg \, s^{-1}}$ (for $d<6.9$~kpc). The \textit{XMM-Newton} luminosity was thus more than an order of magnitude higher than quiescence. 
    \item Comparison with tMSPs: The low-luminosity pulsations and spectral softening are reminiscent of tMSPs in the sub-luminous disk state, where X-ray, optical, and UV pulsations are thought to originate from synchrotron emission at the pulsar wind–disk boundary rather than direct accretion. While this seems a viable option, some key differences remain, such as the lack of a clear bimodality in the X-ray count distribution and the energy dependence of pulsations.
    \item Radio non-detection: We did not detect a radio counterpart during ATCA observations, with a 3$\sigma$ upper limit of 60\,$\mu$Jy\,beam$^{-1}$ at 5.5~GHz on the flux density. Such a limit remains compatible with the radio flux densities observed from AMSPs during outburst and from tMSPs in the sub-luminous disk state, and therefore cannot rule out the presence of a faint or variable radio counterpart. Future time-resolved radio monitoring will also be essential to distinguish between a quiescent AMSP potentially hosting an active radio pulsar and a tMSP-like sub-luminous disk state in IGR~J17511$-$3057.
\end{itemize}

\begin{acknowledgements}
We thank the referee for useful comments. We thank Vito Sguera for the helpful discussions during the outburst.
This work is based on observations obtained with XMM-Newton, an ESA science mission with instruments and contributions directly funded by ESA Member States and NASA. We are grateful to the XMM-Newton director for scheduling ToO observations in Director’s Discretionary Time.
The Australia Telescope Compact Array is part of the Australia Telescope National Facility (\url{https://ror.org/05qajvd42}), which is funded by the Australian Government for operation as a National Facility managed by CSIRO. We acknowledge the Gomeroi people as the Traditional Owners of the ATCA observatory site.

AP acknowledges support by INAF (Research Grant FANS and PULSE-X, PI: Papitto), the Italian Ministry of University and Research (PRIN MUR 2020, Grant 2020BRP57Z, GEMS, PI: Astone), and Fondazione Cariplo/Cassa Depositi e Prestiti (Grant 2023-2560, PI: Papitto). FCZ is supported by a Ramon y Cajal fellowship (grant agreement RYC2021-030888-I). AM and FCZ are also supported by the National Spanish grant PID2023-153099NA-I00.

\end{acknowledgements}

\bibliographystyle{aa} 
\bibliography{bib}

\end{document}